# Piezoelectricity for Nondestructive Testing of Crystal Surfaces

Chandrima Chatterjee[1]

[1]*Department of Physics and Astronomy, University of Mississippi, University, MS, 38677, USA*

A stress is applied at the flat face and the apex of a prismatic piezoelectric crystal. The voltage generated at these points differs in order of magnitude. The result may be used to nondestructively test the uniformity of surfaces of piezoelectric crystals.

**INTRODUCTION**

Piezoelectric crystals find applications in our everyday lives, such as in mobile phones,[1] watches,[2] microphones,[3] and others. Piezoelectricity encompasses the concepts of stress, strain, electric charge, dipole moment that students learn in basic college physics classes,[4,5] thus the topic will be of great interest to them. When we apply stress to a crystal, not only is a strain produced, but also a potential difference is generated between opposite faces of the crystal. This is the direct piezoelectric effect (DPE). Conversely, the indirect piezoelectric effect corresponds to the application of a potential difference between opposite faces to generate strain in the crystal. When no stress is applied, piezoelectric crystals are electrically neutral since a positive charge is cancelled by a negative charge nearby. But when stressed, the atomic structure is deformed and charges are pushed closer together or further apart creating a molecular dipole moment. A net polarization appears and charges accumulate on opposite faces of the crystal. The structure of a piezoelectric crystal is an important criterion based on which it is used for different applications. The aim of this paper is to experimentally study the DPE at different locations in a prismatic piezoelectric crystal. This is because a prism has more than one geometric structure, such as flat faces and a sharp apex. The piezoelectric response from these different macrostructures may be applied in the nondestructive testing of the surfaces of piezoelectric crystals in general.

**EXPERIMENTAL TECHNIQUE**

The sample used in the experiment is shown in Fig. 1. The sample consisted of epitaxial colemanite ($Ca_2B_6O_{11} \cdot 5H_2O$) commercially available at Amargosa Minerals.

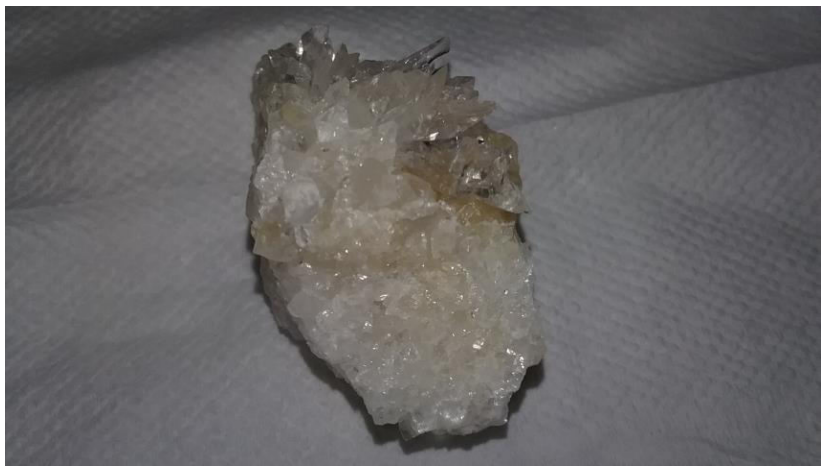

Fig. 1. The cluster of prismatic colemanite crystals is epitaxially grown on a substrate of calcite. The dimensions of the sample are 2.6 cm x 2.3 cm x 1.3 cm. The height of the colemanite prisms varies between 3 mm to 8 mm. The size of the apex of the prisms is less than 10 micron.

The experimental setup to observe the direct piezoelectric effect from colemanite is shown in Fig. 2. Copper electrodes were deposited on opposite faces of the sample. The electrodes were connected to a digital multimeter that reads the voltage. A sharp metal tip of 0.25 mm$^2$ area applied the force. The force was approximately varied between 5000 dyne and 45000 dyne by applying quarter dollar coins on the metal tip. The cluster of crystals did not grow in one direction, but were rather spread out. So the metal tip could land near a prism apex (points 1, 2, 5 and 6 in Fig. 2), or on a prism flat surface (points 3 and 4 in Fig. 2). Data were taken from points close to the apex and the flat surface of long, well-defined prisms. Colemanite is piezoelectric below 270$^o$K.[6] This is because the symmetry of the crystal changes at low temperatures. Therefore the experimental setup was inserted in the freezer unit of a refrigerator and the temperature was set to a minimum. This ensures that the temperature stays below 270$^o$K. As the freezer door is opened to record a reading, a time gap of 15 minutes is given before each new reading, for the temperature to stabilize.

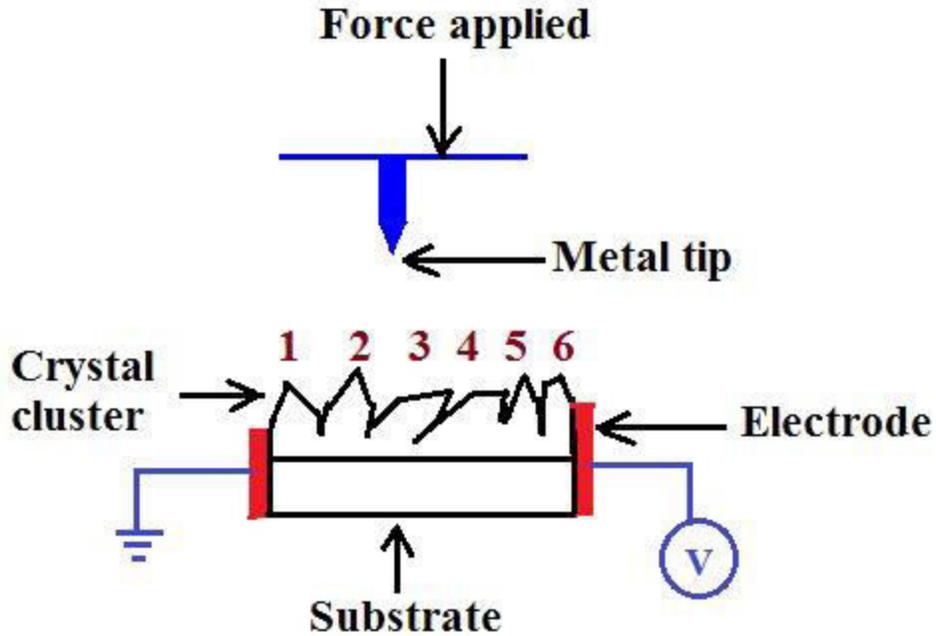

Fig. 2. Schematic diagram of the experimental setup to observe the direct piezoelectric effect in epitaxial colemanite.

**RESULTS AND DISCUSSION**

As a force is applied by placing quarter dollar coins (Q) on the metal tip, a voltage is generated. Fig. 3 shows the variation of the voltage with the applied force when the flat face of the prism is stressed. The voltage increases when in steps of 0.1 mV, as Q increases from 1 to 3 in steps of 1. This is a consequence of the DPE. As the face of the prism is stressed, the charge neutrality is no longer maintained. The face is deformed and a net polarization is developed, which leads to accumulation of charge on the electrodes. The deformation increases with the increase in stress from Q = 1 to Q = 3. This is the reason why the plot is linear in this region. After that, the voltage remains constant until Q = 5. This is because the more the atoms come closer due to an increase in stress, the more the electrons start repelling each other. Thus it becomes more difficult to deform the crystal, and the voltage generated is constant in this region. The voltage increases by 0.1 V when Q is changed from 5 to 6 due to the DPE. After that, the voltage remains constant from Q = 6 to Q = 8. Q was not increased beyond 8 since the metal tip was unsteady beyond this point.

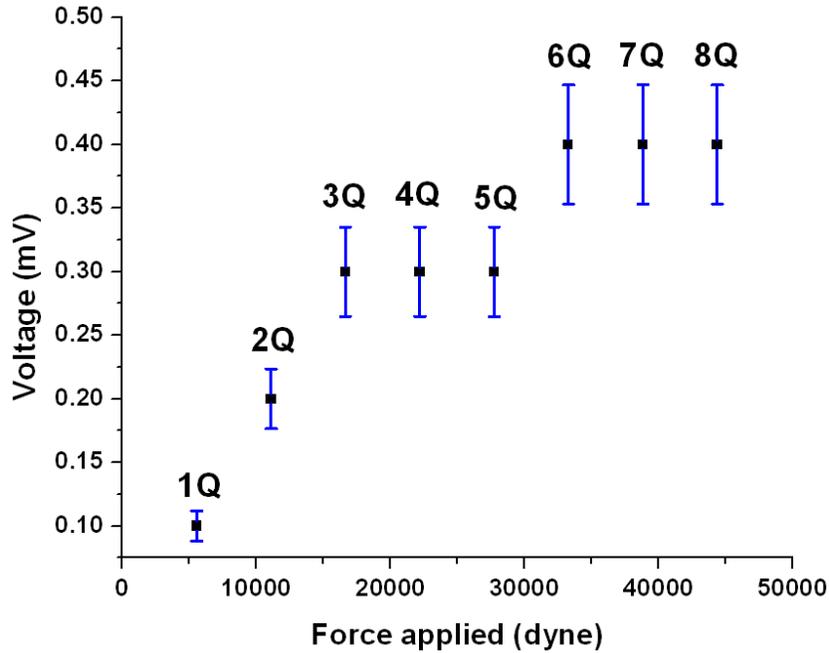

Fig. 3. Variation of voltage with force applied on the face of the prismatic piezoelectric crystal. nQ (where n=1 to n=8) on the plot shows the number of quarters (Q) used to generate the force in dyne. For example, the first point is produced by placing 1 quarter (or 1Q) on the tip.

The apex of the prism is very sharp. It was not possible to balance the metal tip on the apex. So it was balanced at a point slightly lower than the apex. Fig. 4 shows the variation of the voltage with the applied force at a point very close to the apex. It is observed that the voltage generated is equal to 5.2 mV for Q = 1 and Q = 2. At Q =3, the voltage rises to 5.5 mV and then comes down to 5.2 mV at Q = 4. Higher values of Q were not used, as the metal tip could not be kept steady. The voltage generated at the apex is one order of magnitude higher than at the flat surface. This behavior can be explained by the presence of charged defects in the crystal. In addition to the host atoms, a crystal contains point defects such as impurity atoms, interstitial atoms, vacancies and others. The sharp apex of the prism attracts these charged defects, which are localized around it. When points close to the apex are subjected to stress, the charged point defects add to the net polarization. This leads to an increase in the voltage generated. Thus it can be concluded that the voltage depends on the terrain of the point being stressed. Therefore this method can be used to nondestructively test the uniformity of the surfaces of piezoelectric crystals.

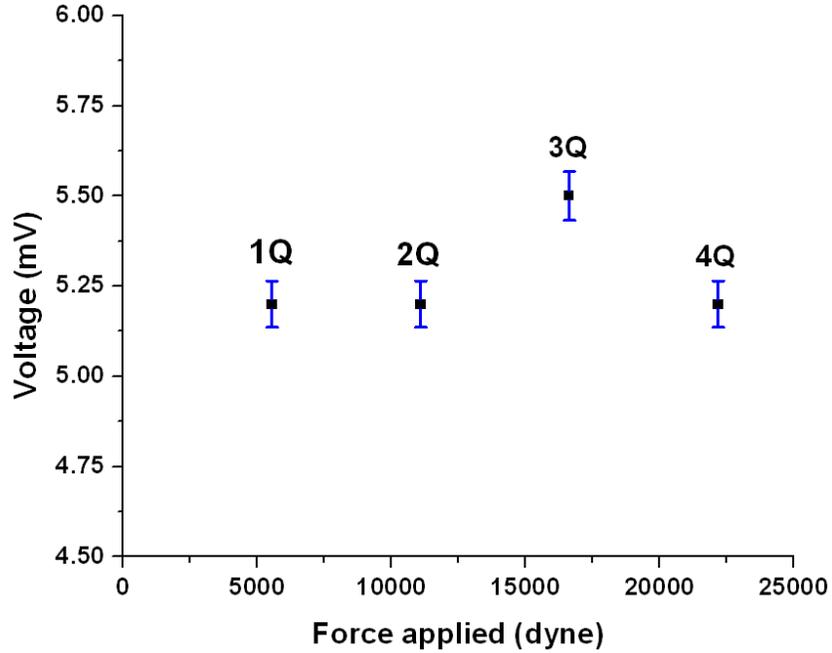

Fig. 4. Variation of voltage with force applied near the apex of the prismatic piezoelectric crystal.

In conclusion, the direct piezoelectric effect was experimentally investigated from different locations of an epitaxial prismatic crystal. When the flat face of the prism is stressed between 5000 dyne and 45000 dyne, the voltage varies between 0.1 mV and 0.4 mV. When a point near the apex of the prism is stressed between 5000 dyne and 25000 dyne, the voltage varies between 5.2 mV and 5.5 mV. It is argued that the increase of voltage near the apex is due to the presence of charged crystal defects. The order of magnitude of the voltage generated depends on the terrain of the point being stressed. The result may be used in the nondestructive testing of the uniformity of the surfaces of piezoelectric crystals in general.